\documentclass[final,3p,times,twocolumn]{elsarticle}
\usepackage{amssymb}
\usepackage{amsmath}
\usepackage{arydshln}
\usepackage{color}
\def\vect#1{\mbox{\boldmath $#1$}}
\def\svect#1{\mbox{\boldmath {\scriptsize $#1$}}}
\def \rmd{\mathrm{d}}
\def \rme{\mathrm{e}}

\newcommand\T{\rule{0pt}{0.6cm}}       % Top strut
\newcommand\B{\rule[-0.4cm]{0pt}{0pt}} % Bottom strut

\journal{Physics Letters A}

\begin{document}

\begin{frontmatter}



\title{Rattleback: a model of how geometric singularity induces dynamic chirality}

\author[label1]{Z. Yoshida}
\ead{yoshida@ppl.k.u-tokyo.ac.jp}

\author[label2]{T. Tokieda}
\ead{tokieda@stanford.edu}

\author[label3]{P.J. Morrison}
\ead{morrison@physics.utexas.edu}

\address[label1]{Department of Advanced Energy, University of Tokyo, Kashiwa, Chiba 277-8561, Japan}
\address[label2]{Department of Mathematics, Stanford University, Stanford CA 94305-2015, USA}
\address[label3]{Department of Physics and Institute for Fusion Studies, University of Texas at Austin, Austin TX 78712-1060, USA}

\begin{abstract}
The rattleback is a boat-shaped top with an asymmetric preference in spin.
Its dynamics can be described by nonlinearly coupled pitching, rolling, and spinning modes.  The chirality, 
designed into the body as a skewed mass distribution, manifests itself in the quicker transition of $+$spin $\rightarrow$ 
pitch $\rightarrow$ $-$spin than that of $-$spin $\rightarrow$ roll $\rightarrow$ 
$+$spin.  
The curious guiding idea of this work is that we can formulate the dynamics 
as if a symmetric body were moving in a chiral space.
By elucidating the duality of matter and space in the Hamiltonian formalism,
we attribute asymmetry to space.
The rattleback is shown to live in the space dictated by the Bianchi type ${\rm VI}_{h < -1}$ (belonging to class B) algebra;
this particular algebra is used here for the first time in a mechanical example.  
The class B algebra has a singularity that separates the space (Poisson manifold) into mirror-asymmetric subspaces,
breaking the time-reversal symmetry of nearby orbits.
\end{abstract}


\end{frontmatter}


\section{Introduction}
\label{sec:introduction}

\begin{figure*}
\centering
\raisebox{3.6cm}{{\large \textbf{a}}}
\includegraphics[scale=0.25]{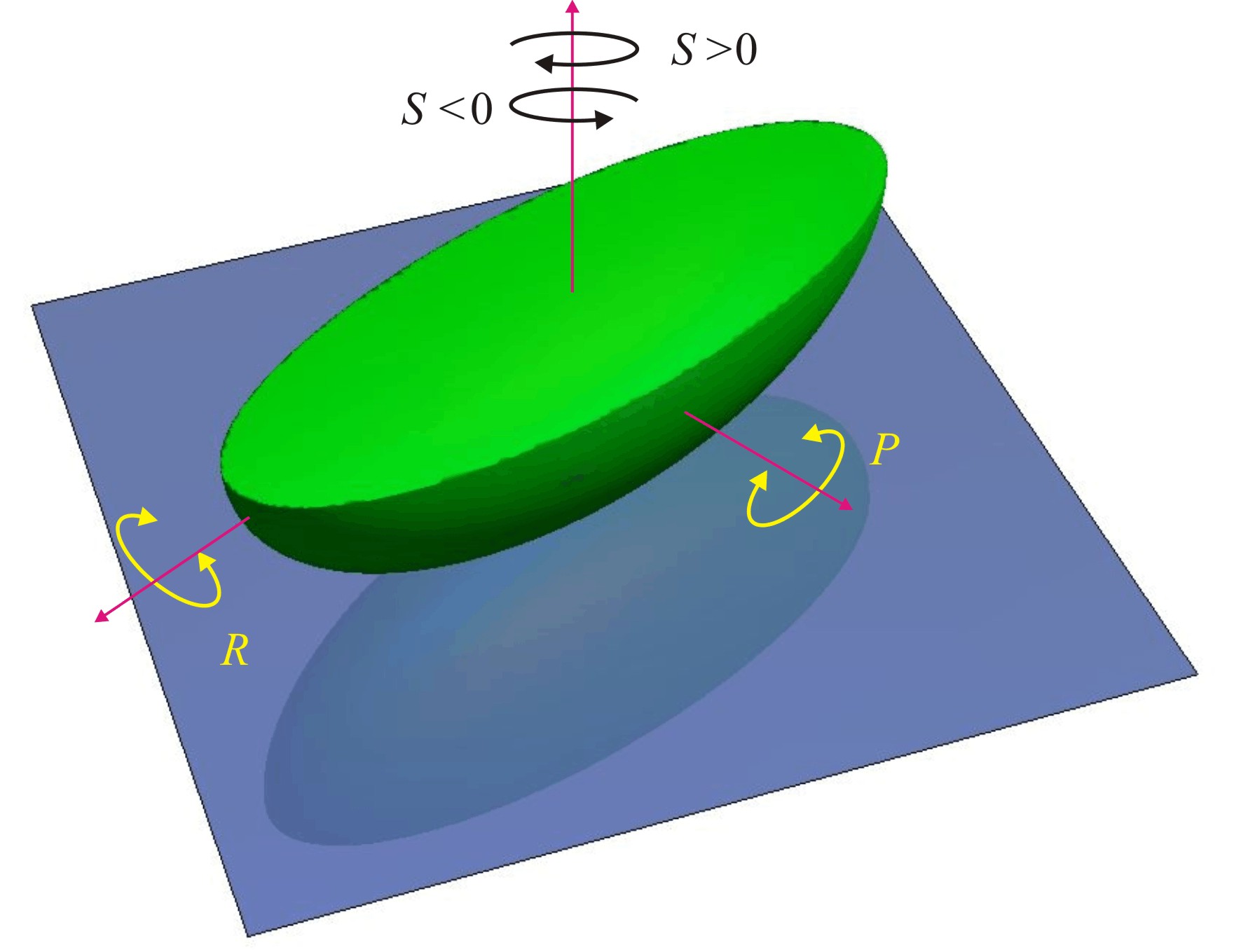}
% \\~\\
~~~
\raisebox{3.6cm}{{\large \textbf{b}}}
\includegraphics[scale=0.7]{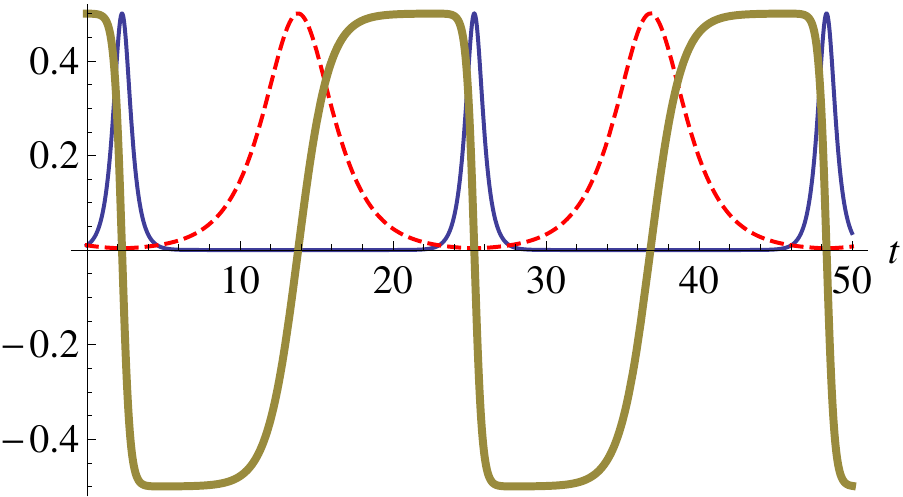}
\caption{
\label{fig:celt0}
(\textbf{a})  A rattleback.  Three modes, \emph{pitch}, \emph{roll}, and \emph{spin},
constitute a nonlinearly coupled dynamics.  Pitch and roll are oscillations, whose intensities are $P$ and $R$.
Spin $S$ takes signs in which the chirality manifests itself:
spinning in the non-preferred sense $S > 0$ induces strong $P$-instability, resulting in a quick reversal,
whereas spinning in the preferred sense $S < 0$ induces weak $R$-instability, resulting in a sluggish reversal.
%%%%%
(\textbf{b})  Typical solution of PRS, performing repeated spin reversals (from Fig.~1 of \cite{Moffatt-Tokieda2008}).
$\lambda=4$, $P(0)=R(0)=0.01$, $S(0)=0.5$.
The blue curve is pitch $P$, the red dotted curve is roll $R$, the
brown thick curve is spin~$S$. 
}
\end{figure*}

The rattleback has amused and bemused people of all walks of life.  For this boat-shaped 
top (Fig.\,\ref{fig:celt0}) to be a rattleback it is sufficient that inertia and geometry be 
misaligned---that the axes of the ellipsoid of inertia be skewed with respect to 
the principal curvature directions of the contact surface.  The chirality in motion, one spin 
more prominent than the opposite spin, points to an 
effect that may occur in complex dynamical systems.  
The minimal model that captures this chiral dynamics is the prototypical 
rattleback system, \emph{PRS\/}~\cite{Moffatt-Tokieda2008}.
We shall see that the non-dissipative version of PRS admits an odd-dimensional, degenerate Hamiltonian formulation.
This is puzzling for two reasons.  i)~A linearized Hamiltonian system has symmetric spectra, 
so should be time-reversible; yet chiral dynamics is not.  ii)~PRS has an extra
conserved quantity besides energy, which hitherto has received no intuitive interpretation.  
The key to the puzzles is a peculiar Lie-algebraic structure behind the Hamiltonian formulation, a so-called Bianchi class B algebra.

This paper is organized as follows.  After reviewing PRS, we write it as a Hamiltonian system and relate it 
to the Bianchi class B type ${\rm VI}_{h < -1}$ algebra.   We visualize the orbits in
the phase space, and remark that this algebra has a singularity, which distorts nearby Casimir
leaves.  In Darboux coordinates (as canonical as manageable in an odd-dimensional space), the system is 
revealed to be a 1-dimensional oscillator in disguise, having an asymmetric potential.  The asymmetry renders 
intuitive the rattleback's chiral behavior; geometrically it comes from the distortion of the Casimir leaves, so ultimately 
from the singularity.  All Bianchi class B algebras have singularities, which hint at where we may look for
further examples of chiral dynamics.

%---------------------------------------------------------------  FIG 1
%----------------------------------------------------------------------

%%%%%%%%%%%%%%%%%%%%%%%%%%%%%%%%%%%%%%%%%%%%%%%%%%%%%%%%%%%%%%%%%%%%%%%%%%%%%%%%%
\section{Prototypical rattleback system}
\label{sec:PRS}

The PRS equation, without dissipation, is
\begin{equation}
\frac{\rmd}{\rmd t} \left( \begin{array}{c}
P \\ R \\ S
\end{array} \right)
=
\left( \begin{array}{c}
\lambda P S \\ - RS \\ R^2 - \lambda P^2
\end{array} \right) .
\label{MT-1}
\end{equation}
Compared with Eq.~5.5 of \cite{Moffatt-Tokieda2008}, we adopted a more felicitous 
notation where $P, R, S$ stand for the {\it pitching}, {\it rolling}, {\it spinning\/} modes of the motion.  
We denote the state vector by $\vect{X} = (P~R~S)^{\rm T} \in {\bf R}^3$.
The parameter $\lambda$ encodes the aspect ratio of the rattleback shape.
Throughout we choose $\lambda>1$.  This means that $P$ corresponds to lengthwise oscillations along the keel 
of the boat, $R$ to sideways oscillations.  When $\lambda = 1$, the rattleback has an umbilic
on the contact surface, so chirality disappears.

Let us examine a number of phenomenological features characteristic of the rattleback.

\subsection{Chiral dynamics}
Figure~\ref{fig:celt0} depicts a typical solution.  
Energy cycles in the order $+S \rightarrow P \rightarrow -S \rightarrow R \rightarrow +S$, where the 
signs $\pm$ distinguish between the two senses of spin.  The chirality manifests 
itself in the transition $+S\rightarrow P\rightarrow -S$ happening quicker than the transition $-S \rightarrow R\rightarrow +S$.
You can think of $-S$ as the sense induced by $P$, of $+S$ as that induced by $R$.  
In terms of the skewness of the mass distribution,
the spin from the long axis of the ellipsoid of inertia to the small principal curvature direction has the $+$ sign

The widespread belief that `one spin is stable and the opposite spin is unstable' is wrong.  
A handy way to check experimentally that in reality both spins are unstable is to start from rest: if we excite pitch by 
tapping on the prow of the boat, the rattleback goes into one spin (this is our $+$); if, however, we excite roll by tapping 
on a side, it spontaneously goes into the opposite spin.  

Both spins are unstable, but the exponents of instability are unequal.  
Linearizing \eqref{MT-1} around the spinning equilibrium $(0~0~S_e)^{\rm T}$ at any constant value  $S_e$,
we find
\begin{equation}
\frac{\rmd}{\rmd t} 
\left( \begin{array}{c}
\Delta P \\ \Delta R \\ \Delta S
\end{array} \right)
=
\left(
\begin{array}{c}
\lambda S_e \Delta P
\\
- S_e \Delta R
\\
0
\end{array} \right).
\label{SeqEom}
\end{equation}
where $\Delta$ denotes perturbations, $P = P_e + \Delta P = \Delta P$, etc.
During $S_e > 0$, $\Delta P$ grows exponentially at a large (quick) rate $\lambda S_e$ while 
$\Delta R$ decays exponentially at a small (sluggish) rate $S_e$.  During $S_e < 0$, 
$\Delta R$ grows at a sluggish rate $|S_e|$ while $\Delta P$ decays at a quick rate 
$\lambda |S_e|$.

Witness the unequal exponents experimentally by placing a rattleback on a vibrating floor.  Though both
pitch and roll get excited, soon pitch dominates and the rattleback ends up with a $-$spin.
We can also shake an ensemble of rattlebacks and create a `chiral gas' \cite{Lubensky}
or a `chiral  metamaterial' \cite{application1,application2,application3}.
% {\tt For the record, this is not the gas I was suggesting.  Mine is a real kinetic theory that has a probability density, mean field or Liouville, with the PRS phase space.  The one of this reference [2] is highly phenomenological and not derived from the primitive PRS dynamics. }
% \textcolor{blue}{
Such systems have a chance of motivating innovation of various technologies. For example, chiral `particles' may be used for energy harvesting, in a manner similar to the mechanism of automatic wristwatch winding, where they convert ambient thermal fluctuations to some lower-entropy energy. The following formulation and analysis provides a basic picture for designing such devices.
% }

It can be shown that given the above phenomenological features, subject to the hypothesis
that the model be 1st-order and quadratic in $P, R, S$, the form of the equation of motion is essentially unique.
Thus PRS is actually the minimal model of the rattleback.

\subsection{Conservation laws}
  In experiments we seldom see the $-$spin go into 
roll and reverse, because dissipation (friction on the floor) tends to kill the motion before 
this weak instability kicks in.  The $+$spin, whose instability is strong, goes into pitch vigorously and reverses
quickly.  
If dissipation is absent or not too severe, the rattleback keeps reversing back and forth.  
Indeed, thanks to the following 2 conserved quantities~\cite{Moffatt-Tokieda2008},
PRS is integrable and the orbits are periodic:
\begin{eqnarray}
H(\vect{X})
&=&  \frac{1}{2}\left( P^2 + R^2 + S^2 \right),
\label{MT-H}
\\
C(\vect{X}) &=& P R^\lambda .
\label{Casimir}
\end{eqnarray}
The intersection of a level surface (sphere) of $H$ and of a level surface (leaf) of $C$ delineates 
the orbit in the 3-dimensional phase space ${\bf R}^3$;
see Fig.~\ref{fig:leaves} (the picture is extended to $P < 0, R < 0$).

   The orbit may reduce to an equilibrium.  In Fig.~\ref{fig:leaves} we notice, for each value of $H$, 4 points at which 
the energy sphere is tangent to a Casimir leaf.  Physically we mix just the right balance
of pitch and roll, $R/P = \sqrt{\lambda}$, for a given energy so that the rattleback rocks with zero spin.  This is expected: pure pitch induces $S < 0$, pure roll induces $S > 0$, so at some azimuth in-between we must be able to induce $S = 0$.  These rocking equilibria 
(physically 2) are stable, because again by Fig.~\ref{fig:leaves}, if we perturb the values of $H$ or $C$, 
periodic orbits appear nearby.  
%(Stability is verified also by the `energy-Casimir method', i.e.\ Lagrange-multiplier extremization with the Casimir as 
%a constraint, plus stability analysis using perturbations admissible for the constraint.)

\subsection{Theoretical problems}
    However, the conclusion of the eigen-modal analysis is puzzling when viewed against the general
Hamiltonian theory (as about to be shown, PRS is Hamiltonian).  Every linear Hamiltonian system 
has time-reversible symmetric spectra (Krein's theorem): if $\mu \in {\bf C}$ is an eigenvalue, so are 
$-\mu, \mu^*, -\mu^*$.  The unequal exponents $\lambda S_e \neq S_e$ violate this symmetry.
The culprit for this violation is that our Hamiltonian system is \emph{singular\/} (meaning
`not regular', rather than `infinite') at $(0~0~S_e)^{\rm T}$.
The original nonlinear system \eqref{MT-1} has a parity-time symmetry 
(invariant under $t\mapsto-t$ and $(P~R~S)^{\rm T} \mapsto (P~R~-S)^{\rm T}$).

Another puzzling issue is that in \eqref{MT-H} and \eqref{Casimir}, $H$ is the energy, but the extra conserved
quantity $C$ seems to come from nowhere (its spotting in \cite{Moffatt-Tokieda2008} was serendipitous).  

Inquiry into these questions reveals an interesting, and possibly general, mathematical structure
that imparts chirality to a dynamical system.

%%%%%%%%%%%%%%%%%%%%%%%%%%%%%%%%%%%%%%%%%%%%%%%%%%%%%%%%%%%%%%%%%%%%%%%%%%%%%%%%%%%%%%%%%%%%%%%
\section{Analysis}
\label{sec:Hamiltonian}
   We shall now see that the geometry of the phase space and the underlying Lie-Poisson algebra allow us to 
go some way toward a more fundamental explanation of the dynamic chirality and an interpretation of $C$, as follows.

%---------------------------------------------------------------  FIG 2
\begin{figure}[tb]
\begin{center}
\includegraphics[scale=0.4]{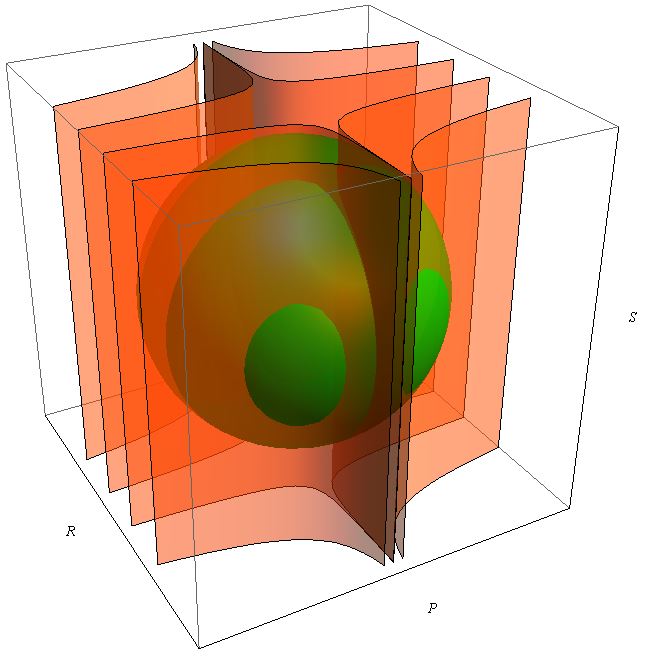}
\caption{
\label{fig:leaves}
Orbits are the intersections of an energy sphere $H=\frac{1}{2}(P^2+R^2+S^2) =$ const 
(green) and a Casimir leaf $C=P R^\lambda =$ const (red).
The sphere $H = 1$ and the leaves $C=-1, -0.01, +0.01, +1$ are depicted. 
The aspect ratio is $\lambda=4$.
}
\end{center}
\end{figure}
%----------------------------------------------------------------------

\subsection{Hamiltonian structure}
Despite the odd-dimensionality of the phase space ${\bf R}^3$, it proves possible to write the PRS~\eqref{MT-1} 
as a Hamiltonian system
whose Poisson bracket on the phase-space functions (observables) realizes some known Lie algebra.  Such 
a bracket is known as a 
\emph{Lie-Poisson bracket}~\cite{Morrison1998}.
% \textcolor{blue}{
There is a large literature on the Hamiltonian structure of generalized tops and associated Lie algebra constructions;
see e.g.\ \cite{neishtdat,thiffeault,tsiganov} and references therein.
% }

Denoting by $\langle\, , \rangle$ the usual inner product on ${\bf R}^3$,
we define a bracket on observables
\begin{equation}
\{ F, G\}_J = \langle \partial_{\svect{X}} F, J \partial_{\svect{X}} G \rangle 
\label{Poisson}
\end{equation}
with a Poisson matrix  
\begin{equation}
J  = \left( \begin{array}{ccc}
0  &  0  &  \lambda P
\\
0  &  0  &  -R
\\
-\lambda P & R  & 0
\end{array} \right).
\label{MT-J}
\end{equation}
This bracket is Lie-Poisson, realizing the \emph{Bianchi type} $\mathrm{VI}_h$ ($h= -\lambda$)
\emph{Lie algebra\/} on the space of observables (see Table\,\ref{table:Bianchi-Casimirs} and  Appendix A).  
The unfamiliarity of this algebra 
attests to the strange behavior of the rattleback and vice versa.
We have $\textrm{rank}\, J = 2$ except along the singular locus,
the $S$-axis $P = R = 0$, where $\textrm{rank}\,J$ drops to $0$.
We also have $J \partial_{\svect{X}}C =0$ so $\{C, G\}_J =0$ for every observable $G$, i.e.\
$C$ is a \emph{Casimir}.  In Fig.\,\ref{fig:leaves} the Casimir leaves are colored red.  
With $H$ as the Hamiltonian,
\eqref{MT-1} becomes Hamilton's equation
\begin{equation}
\frac{\rmd }{\rmd t} \vect{X} = \{ \vect{X}, H \}_J \,.
\label{Hamilton_eq}
\end{equation}
For every orbit, there exists a leaf on which that orbit lies entirely.  A Casimir leaf is the
effective phase space for that value of $C$.

%    Be it noted in passing that another interpretation of $C$ is available.  Such a degenerate Hamiltonian
% system embeds, on `unfreezing the Casimir', into a 4-dimensional non-degenerate system, 
% in which $C$ emerges as an adiabatic invariant conjugate to a tiny-scale angle variable, 
% the augmented 4th dimension; see \cite{Yoshida-Morrison2014,Yoshida-Morrrison2016}.  
% In our case Eq.\ref{Poisson})--Eq.\ref{Hamilton_eq}), the
% projection of the 4-dimensional dynamics on ${\bf R}^3$, the space of $(P~R~S)^{\rm T}$, is chaotic.

\subsection{Dual formalism}
% \subsection*{Deal Formalism Switching between Hamiltonian and Casimir}
  In our formalism (after rescaling), the Hamiltonian $H$ of \eqref{MT-H} was symmetric, the 
Casimir of \eqref{Casimir} was asymmetric: it is as if a symmetric body were moving in an asymmetric
phase-space.  But an orbit is just the intersection of two level surfaces and does not know which is
Hamiltonian and which is Casimir.  Hence, there ought to be a dual Lie-Poisson formalism that exchanges 
the roles of $H$ and $C$, as if an asymmetric body were moving in a symmetric phase space.  Though this 
sounds pleasant---after all, the rattleback is an asymmetric body---the calculations are less so.
The Poisson matrix
\begin{equation}
K  = \left( \begin{array}{ccc}
0   &  R^{1-\lambda}S  &  -R^{2-\lambda}
\\
- R^{1-\lambda}S  &  0  &  PR^{1-\lambda}
\\
R^{2-\lambda} & -PR^{1-\lambda}  & 0
\end{array} \right)
\label{MT-K}
\end{equation}
is found by the requirement that \eqref{Hamilton_eq} with $C$ replacing $H$ should recover 
the PRS~\eqref{MT-1}.  Since $K$ is nonlinear in $\vect{X}$, the bracket  
$\{ F, G \}_K = \langle \partial_{\svect{X}} F, K \partial_{\svect{X}} G\rangle$
is not Lie-Poisson 
(unless $\lambda=1$).
But a change of coordinates 
\[
\vect{X} \mapsto
\vect{Y} 
= 
 \left( \begin{array}{c}
\sqrt{P^2+S^2}\cos [ \, R^{1-\lambda}\arctan(S/P)\, ]  \\ R \\ \sqrt{P^2+S^2}\sin [ \, R^{1-\lambda}\arctan(S/P)\, ]
\end{array} \right)
\]
turns $K$ into an $\mathfrak{so}(3)$-matrix
\begin{equation}
L = \left( \begin{array}{ccc}
0  &   Y_3 &  -Y_2
\\
-Y_3 &  0  &  Y_1
\\
Y_2 &  -Y_1  & 0
\end{array} \right),
\label{so3}
\end{equation}
and we have a Lie-Poisson bracket 
$\{ F, G \}_L = \langle \partial_{\svect{Y}} F, L \partial_{\svect{Y}} G\rangle$, type IX.
The Hamiltonian is
\begin{equation}
C(\vect{Y}) = \pm \frac{Y_2^\lambda \sqrt{Y_1^2 + Y_3^2}}{\cos [ \, Y_2^{\lambda-1}\arctan(Y_3/Y_1)\, ] }.
\label{second-Hamiltonian}
\end{equation}

\subsection{Casimir leaf ---skewed effective space}
\label{sec:Casimir}
   Return to the Lie-Poisson bracket $\{ F, G\}_J$ of \eqref{Poisson}.  The natural thing to do is to bring it
into a normal form.  This is achieved by Darboux's theorem: $J$  is equivalent to
\begin{equation}
J_{D}  = \left( \begin{array}{ccc}
0  &   1 &  0
\\
-1 &  0  &  0
\\
% \hdashline
0  &  0  & 0
\end{array} \right) 
\label{normal-J}
\end{equation}
under a change of coordinates in which the  Casimir serves as one coordinate and the other two coordinates
are a canonically conjugate pair that parametrize the Casimir leaves.  Explicitly,
\begin{equation}
\vect{X}
\mapsto 
\vect{Z} 
= \left( \begin{array}{c}
-\log R  \\ S \\ P R^\lambda \end{array} \right)
 \label{trans}
\end{equation}
defines the normal form of the (still degenerate) Poisson bracket
$\{F, G\}_{J_D} 
= \langle \partial_{\svect{Z}} F, J_{D} \partial_{\svect{Z}} G \rangle
= \{F, G\}_J$.
In these almost canonical \emph{Darboux coordinates\/} $\vect{Z}$ the Hamiltonian of \eqref{MT-H} takes the form
\begin{equation}
H(\vect{Z})  =   \frac{1}{2}{Z}_2^2 + U_{\lambda,C}({Z}_1) 
\label{MT-H(zeta)'}
\end{equation}
with
\begin{equation}
 U_{\lambda,C}({Z}_1) = \frac{1}{2} \left( \rme^{-2{Z}_1} + C^2\rme^{2\lambda{Z}_1} \right).
\label{potential}
\end{equation}
It is sensible to regard ${Z}_1$ as the position and  ${Z}_2$ as the velocity 
of an oscillator, because ${\rm d}Z_1/{{\rm d} t}  = Z_2$ by the equation of motion.
Then the first and second terms on the right of \eqref{MT-H(zeta)'} are kinetic and potential energies.  
$\lambda$ and $C$ are constants, 
the value of the Casimir $C={Z}_3$ being fixed by the initial condition.

%---------------------------------------------------------------  FIG 3
\begin{figure}[tb]
\begin{center}
\includegraphics[scale=0.7]{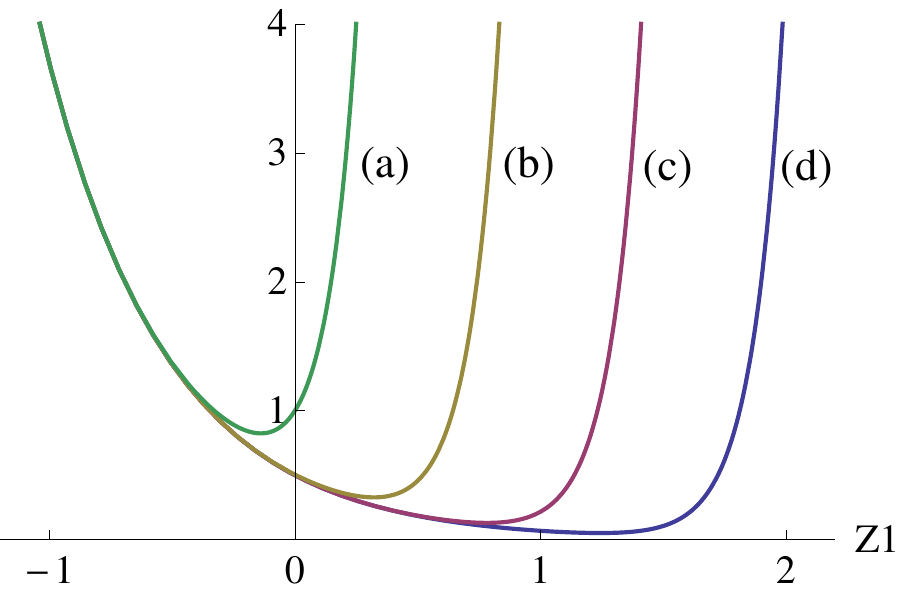}
\caption{
\label{fig:potential}
The asymmetric potential 
$U_{\lambda,C}({Z}_1) = \frac{1}{2} (\rme^{-2{Z}_1} + C^2\rme^{2\lambda{Z}_1})$ of the canonized equation 
of motion;
$\lambda=4$, (a) $C=1$, (b) $C=0.1$, (c) $C=0.01$, (d) $C=0.001$.
}
\end{center}
\end{figure}
%----------------------------------------------------------------------

  It is now easy to read off the rattleback's behavior from the asymmetric shape of the potential energy 
$U_{\lambda,C}({Z}_1)$.  As graphed in Fig.~\ref{fig:potential}, for $\lambda>1$ and $C\neq0$ the potential
has gentle slopes toward negative $Z_1$ and steep cliffs toward positive ${Z}_1$.
Suppose we send the rattleback toward positive $Z_1$.  It runs up a steep cliff and sharply reverses the spin, 
turning ${Z}_2 = S$ from positive to negative (cf.\,Fig.\,\ref{fig:celt0}). 
As it moves toward negative $Z_1$, it trots up a gentle slope and eventually turns from negative to
positive $Z_2$, but the reversal is not so sharp.
The bottom of the potential, $Z_1 = \log (C^2\lambda)^{- 1/(2\lambda + 2)}$, 
is the rocking equilibrium.  When $\lambda <1$, the asymmetry is mirrored, the negative-to-positive 
turn being sharper than the positive-to-negative turn.  The value of $C$ has no bearing on chirality.

   Ultimately the asymmetry of $U_{\lambda,C}({Z}_1)$ is imputable to the asymmetry of the Casimir 
$P R^\lambda$  against the background of the 
Hamiltonian $\frac{1}{2} (P^2 + R^2 + S^2)$ symmetric in the 3 variables.  The Casimir leaves in Fig.~\ref{fig:leaves} are
distorted near the $S$-axis, the singularity.  The presence of a singularity and the asymmetric distortion 
of Casimir leaves near the singularity are common to all Bianchi class B algebras, of which Bianchi type 
${\rm VI}_{h < -1}$, realized by our Lie-Poisson bracket $\{ F, G\}_J$, was an instance.
   
% \textcolor{blue}{
We have solved the puzzles given in section 2.3: the breaking of the time-reversible symmetry of spectra is due to the singularity of the Poisson matrix, which prevents the application of Krein's theorem.  The unbalanced growth rates are for orbits near the equilibrium point that lives just at the singularity (where the rank of the Poisson matrix drops to zero, so every point of the singularity is an equilibrium point, independently of any particular Hamiltonian).  Unlike the perturbations around usual equilibrium points in regular Hamiltonian systems, the perturbation around the singularity affects the Poisson matrix itself, resulting in strange non-Hamiltonian spectra.
% }

% \textcolor{blue}{
We have identified the invariant $C$ as the Casimir of the governing Poisson algebra.  In the next section, we will put Casimirs into perspective.
% }

\section{Variety of chiral systems: Bianchi classification}
\label{sec:Bianchi}

\begin{figure*}[tb]
\begin{center}
~ \\
\includegraphics[scale=0.5]{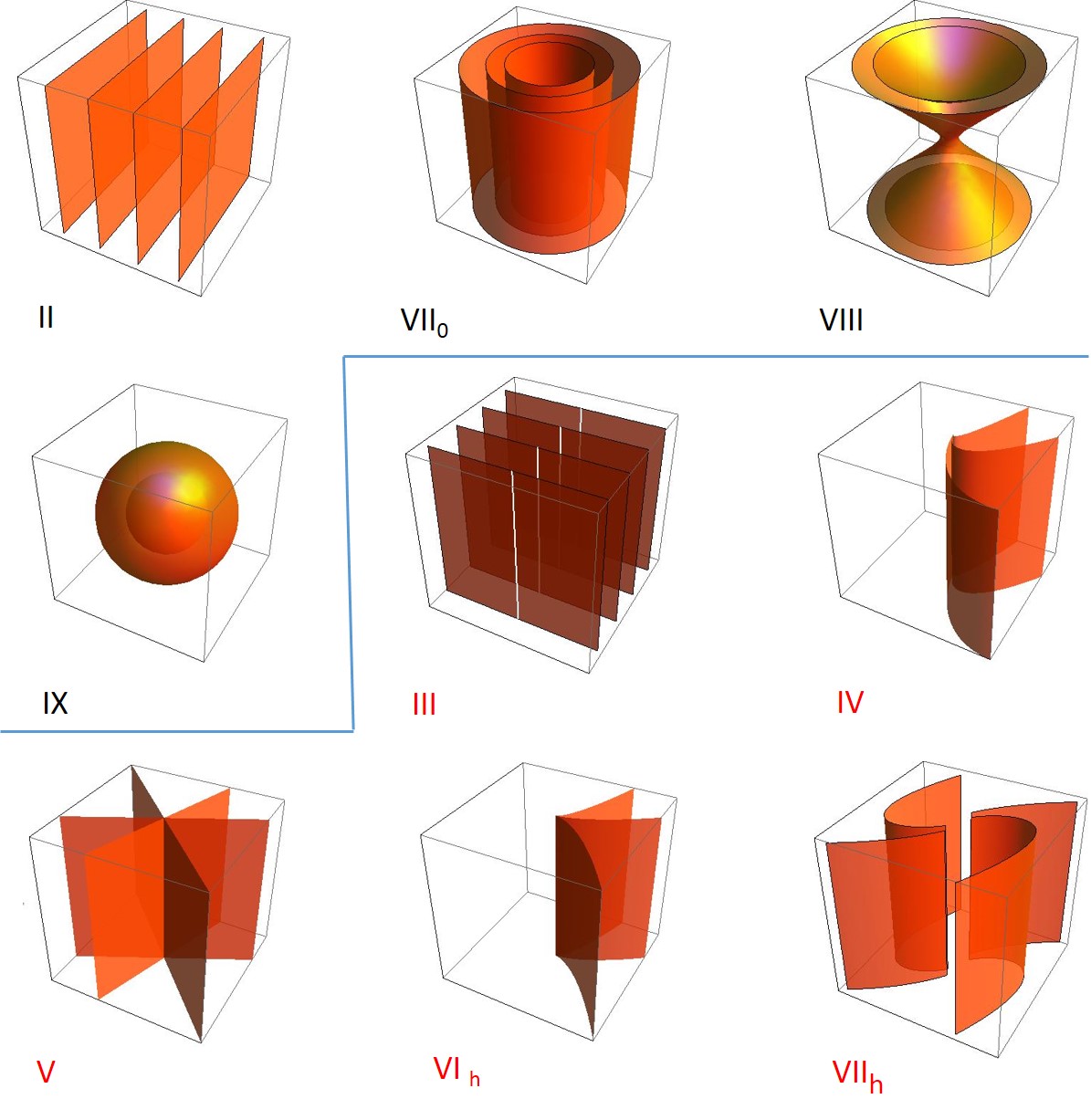}
\caption{
\label{fig:Bianchi_leaves}
Bianchi algebras foliated by Casimir leaves.
Type I algebra is commutative, so the Lie-Poisson bracket is trivial. 
Class~A,  composed of types $\mathrm{I}, \mathrm{II}, \mathrm{VI}_{-1}, \mathrm{VII}_0, \mathrm{VIII}, \mathrm{IX}$,  have regular leaves, whereas class B, composed of types $\mathrm{III}, \mathrm{IV}, \mathrm{V}, \mathrm{VI}_{h\neq-1}, \mathrm{VII}_{h\neq 0}$, have singularities.  
Types $\mathrm{II}$ and $\mathrm{III}$ look alike, but the leaves in type III are singular along the vertical white lines.  
The type $\mathrm{VI}_h$ leaves for $h<0$ were already in Fig.\,\ref{fig:leaves};  the leaves here are for $h=1.5$.   
The pictures of the leaves of type $\mathrm{VII}_h$ ($h=1.5$) are cut off near the singularity $X_1=X_2=0$.  
Type $\mathrm{II}$ is the Heisenberg algebra.  Types $\mathrm{IX}$ $\mathfrak{so}(3)$ and
$\mathrm{VIII}$ $\mathfrak{so}(2,1)$ govern the free rigid body and the Kida vortex.  
We have discovered that type  $\mathrm{VI}_{h < -1}$ governs the prototypical rattleback system (PRS). 
}
\end{center}
\end{figure*}

   Figure \ref{fig:Bianchi_leaves} and Table \ref{table:Bianchi-Casimirs} provide a complete list of 3-dimensional Lie-Poisson systems.
The real 3-dimensional Lie algebras are classified according to the scheme for
Bianchi cosmologies (e.g.\ \cite{Wald}).  There are two classes: class A composed of 
 types $\mathrm{I}, \mathrm{II}, \mathrm{VI}_{-1}, \mathrm{VII}_0, \mathrm{VIII}, \mathrm{IX}$, 
 and class B composed of types $\mathrm{III}, \mathrm{IV}, \mathrm{V}, \mathrm{VI}_{h\neq-1}, \mathrm{VII}_{h\neq 0}$. 
Fig.\,\ref{fig:Bianchi_leaves} depicts their Casimir leaves.
All these 3-dimensional systems possess 2 conserved quantities and hence are integrable:  
the Hamiltonian, plus a Casimir that spans the kernel of the degenerate bracket.   

   In class A, the most elementary instance is type II, Heisenberg algebra:
the intersection circles of an energy sphere and a flat Casimir leaf are the orbits of a harmonic oscillator.
For the free rigid body, type IX $\mathfrak{so}(3)$-algebra, the Casimir leaves are the angular momentum spheres:
their intersections with an energy ellipsoid give the `tennis racket theorem'.
For the equation of the Kida vortex in fluid mechanics, type $\mathrm{VIII}$ $\mathfrak{so}(2,1)$-algebra,
the Casimir hyperboloids  and an energy surface intersect in rotational, librational, or unstable orbits of the 
patch\,\cite{Flierl}.  The leaves for class~A, being quadrics, are regular 
except at the zero set (e.g.\ at  the center of the spheres in type $\mathrm{IX}$).    

In contrast, every system that realizes a class B algebra has a singularity and exhibits chirality.
It is remarkable that PRS realizes a class B algebra, type $\mathrm{VI}_{h < -1}$.  It is the first time 
any class B algebra appears in a mechanical example.
% {\tt Same comment as that of the Abstract.} .  
Moreover, in all the other
class B systems, every Casimir leaf is attached to the singularity, leading orbits into asymptotic regimes.
But in type $\mathrm{VI}_{h < -1}$, the singularity is disjoint from the leaves.
This enables the PRS orbits to be closed curves on the leaf, producing periodic rattling and reversals.

%---------------------------------------------------------------  TABLE 1
\begin{table}[tb]
\caption{3-dimensional Lie-Poisson algebras (Bianchi classification).
To avoid redundancy, for type $\mathrm{IV}_h$ we impose $h\neq 0,1$.   The Casimir of type $\mathrm{VII}_{h\neq0}$ 
requires further classification:
$|h| > 2$ gives $C_{\mathrm{VII}_{h\neq 0}}=\lambda_- \log(-\lambda_- X_1 -X_2) - \lambda_+\log (\lambda_+ X_1 +X_2)$;
$h = \pm 2$ gives
$C_{\mathrm{VII}_{h\neq 0}}= \frac{\pm X_2 }{X_1 \mp X_2}+\log(X_1\mp X_2)$;
$-2 < h < 2$ gives, putting $a = -h/2$ and $\omega=\sqrt{-h^2/4}$ (i.e.\ $\lambda_\pm = a\pm i\omega$),
$C_{\mathrm{VII}_{h\neq 0}}= 2 a \arctan \frac{aX_1 + X_2}{\omega X_1} -
\omega \log [ (aX_1 + X_2)^2 + (\omega X_1)^2 ] $.
}
\begin{center}
{\scriptsize
\begin{tabular}{l|l|l}
\hline
Type & Poisson matrix  & Casimir
\\ \hline
I \T\B &
${\, 
% J_{\mathrm{I}}=
\left( \begin{array}{ccc}
\ 0 & \ 0 & \ 0 \\
\ 0 & \ 0 & \ 0 \\
\ 0 & \ 0 & \ 0 \end{array} \right) }$ &
${\,  
% C_{\mathrm{I}}=
\left\{ \begin{array}{c}
X_1 \\ X_2 \\ X_3
\end{array} \right.} $
\\ \hline
II  \T\B & 
${\, 
% J_{\mathrm{II}}=
\left( \begin{array}{ccc}
0 & 0 & 0 \\
0 & 0 & X_1 \\
0 & -X_1 & 0 \end{array} \right) }$ &
$
% C_{\mathrm{II}}= 
X_1 $
\\ \hline
III  \T\B &
${\, 
% J_{\mathrm{III}}=
\left( \begin{array}{ccc}
0 & 0 & \ X_1 \\
0 & 0 & \ 0 \\
-X_1 & 0& \ 0 \end{array} \right) } $ &
$
% C_{\mathrm{III}}= 
X_2$

\\ \hline
IV \T\B  &
${\, 
% J_{\mathrm{IV}}=
\left( \begin{array}{ccc}
\ 0 & \ 0 &  \ X_1 \\
\ 0 & \ 0 & X_1+X_2 \\
-X_1 & \ - X_1-X_2&\   0 \end{array} \right) } $ &
$
% C_{\mathrm{IV}}= 
\frac{X_2}{ X_1}  - \log{X_1}$

\\ \hline
V \T\B &
${\, 
% J_{\mathrm{V}}=
\left( \begin{array}{ccc}
\ 0 & \ 0 &  \ X_1 \\
\ 0 & \ 0 & X_2 \\
-X_1 & \ -X_2&\   0 \end{array} \right) } $ &
$
% C_{\mathrm{V}}= 
\frac{X_2}{ X_1} $

\\ \hline
$\mathrm{VI}_{-1}$\T\B  &
${\, 
% J_{\mathrm{VI}_{-1}}=
\left( \begin{array}{ccc}
0 & 0 & X_1 \\
0 & 0 & -X_2 \\
-X_1 & X_2 & 0 \end{array} \right) } $ &
$
% C_{\mathrm{VI}_{-1}}= 
X_1 X_2 $

\\ \hline
$\mathrm{VI}_{h\neq -1}$ \T\B  &
${\, 
% J_{\mathrm{VI}_{h\neq 0}}=
\left( \begin{array}{ccc}
0 & 0 & X_1 \\
0 & 0 & hX_2 \\
-X_1 & -hX_2 & 0 \end{array} \right) } $ &
$
% C_{\mathrm{VI}_{h\neq 0}}=
 \frac{X_2}{X_1^h}  $

\\ \hline
$\mathrm{VII}_{0}$ \T\B  &
${\, 
% J_{\mathrm{VII}_{0}}=
\left( \begin{array}{ccc}
0 & 0 & X_2 \\
0 & 0 & -X_1 \\
-X_2 & X_1 & 0 \end{array} \right) }$ &
$
% C_{\mathrm{VII}_{0}}= 
X_1^2 + X_2^2$

\\ \hline
$\mathrm{VII}_{h\neq0}$ \T\B &
${\, 
% J_{\mathrm{VII}_{h\neq 0}}=
\left( \begin{array}{ccc}
0 & 0 & \ X_2 \\
0 & 0 & \ -X_1 + hX_2 \\
-X_2 & \  X_1-hX_2 &\  0 \end{array} \right) } $ &
$
%  C_{\mathrm{VII}_{h\neq 0}}=
 G(X_1,X_2,X_3)  $

\\ \hline
$\mathrm{VIII}$ \T\B  &
${\, 
% J_{\mathrm{VIII}}=
\left( \begin{array}{ccc}
0 & X_3 & X_2 \\
-X_3 & 0 & -X_1 \\
-X_2 & X_1 & 0 \end{array} \right) }$ &
$
% C_{\mathrm{VIII}}= 
X_1^2 + X_2^2  - X_3^2$

\\ \hline
$\mathrm{IX}$ \T\B &
${\, 
% J_{\mathrm{IX}}=
\left( \begin{array}{ccc}
0 & X_3 & -X_2 \\
-X_3 & 0 & X_1 \\
X_2 & -X_1 & 0 \end{array} \right) }$ &
$
% C_{\mathrm{IX}}= 
X_1^2 + X_2^2  + X_3^2$

\\ \hline
\end{tabular}
}
\end{center}
\label{table:Bianchi-Casimirs}
\end{table}
%----------------------------------------------------------------------------

%---------------------------------------------------------------  FIG 4

%%%%%%%%%%%%%%%%%%%%%%%%%%%%%%%%%%%%%%%%%%%%%%%%%%%%%%%%%%%%%%%%%%%%%%%
\section{Chaotic rattleback}
\label{sec:phantom}

Integrable Hamiltonian systems like PRS are structurally unstable.  Here we perturb this system in a natural way by appending to its canonical Hamiltonian form an additional degree of freedom.  This allows us to investigate the ensuing chaos of PRS. 

%%%%%%%%%%%%%%%%%%%%%%%%%%%%%%
%%%%%%%%%%%%%%%%%%%%%%%%%%%%%%
% \subsection{Embedding into 4-dimensional phase space}
% \label{subsec:embedding}

We embed the phase space ${\bf R}^3$ into a 4-dimensional phase space ${\bf R}^4$,
and construct an extended Poisson algebra by extending ${J}_{D}$ to a cosymplectic matrix $J_C$.  Adjoining a new coordinate ${Z}_4\in{\bf R}$, let
\[
\widetilde{\vect{Z}} = \left( \begin{array}{c}
\widetilde{{Z}}_1 \\ \widetilde{{Z}}_2 \\ \widetilde{{Z}}_3 \\ \widetilde{{Z}}_4
\end{array} \right) 
=
\left( \begin{array}{c}
{Z}_1 \\ {Z}_2 \\ {Z}_3 \\ {Z}_4
\end{array} \right) 
=
\left( \begin{array}{c}
-\log R \\ S \\ PR^\lambda \\ {Z}_4
\end{array} \right) 
 \in \mathbf{R}^4.
\]
The idea of \emph{canonical extension\/} \cite{Yoshida-Morrison2014,Yoshida-Morrison2016} of ${J}_{D}$ is simple:
let
\begin{equation}
% \widetilde{J}_D = 
J_C=
% \left( \begin{array}{cccc}
\left( \begin{array}{ccc:c}
0 &  1 & 0 & 0 \\
-1 & 0 &0 & 0 \\
0 & 0 & 0 &1 \\
\hdashline
0 & 0 & -1&  0
\end{array} \right) \,,
\label{4D_Canonical-0}
\end{equation}
where the  kernel of ${J}_{D}$ (the 33 zero corner) gets inflated to a symplectic cell
in $J_C$; the extended Poisson matrix $J_C$ is the  $4\times4$ cosymplectic matrix,
which defines the Poisson algebra % $C^\infty_{\{~,~\}_{\widetilde{J}_c}}(\widetilde{\Omega})$ 
with the canonical bracket
(denoting by $\langle~,~\rangle_{\mathbf{R}^4}$ the inner product on $\mathbf{R}^4$)
\begin{equation}
\{ F, G \}_{J_C}  = \langle \, \partial_{\tilde{\svect{Z}}} F, \, 
J_C \,\partial_{\tilde{\svect{Z}}} G \, \rangle_{\mathbf{R}^4}.
%= \sum_{j =1}^4 \sum_{j=1}^4 \tilde{\mathcal{J}}_{jk} \frac{\partial F}{\partial {\tilde{{Z}}}_j}  \frac{\partial G}{\partial {\tilde{{Z}}}_k} ,
\label{canonized_Poisson}
\end{equation}

As long as the Hamiltonian $H$ does not depend explicitly on the new variable ${Z}_4$,
the dynamics on the submanifold ${\bf R}^3$ is the same as the original dynamics. When this is the case, 
we call ${Z}_4$  a \emph{phantom\/} variable.
Now the invariance of $C$, which was a hallmark of the degeneracy of ${J}$,
has been removed: $J_C$ being  canonical
on the extended phase space $\mathbf{R}^4$ has no Casimir invariant.
Instead, $C$ is a first integral coming from the symmetry $\partial_{{Z}_4} H=0$ via Noether's theorem.

However, we can \emph{unfreeze\/} $C$ by perturbing  $H$ with a term containing  the new variable ${Z}_4$;  in which case, 
 $C={Z}_3$ becomes dynamical and ${Z}_4$ becomes an \emph{actual\/} variable.
Physically, we may interpret a Casimir as an \emph{adiabatic invariant\/} associated with
an ignorable, small-scale angle variable~\cite{Yoshida-Morrison2014,Yoshida-Morrison2016}; upon 
adjoining ${Z}_4$ to $H$  the angle variable materializes from phantom to actual.

%---------------------------------------------------------------  FIG 5
\begin{figure}[tb]
\begin{center}
\raisebox{4.3cm}{\textbf{a}}~~
\includegraphics[scale=0.4]{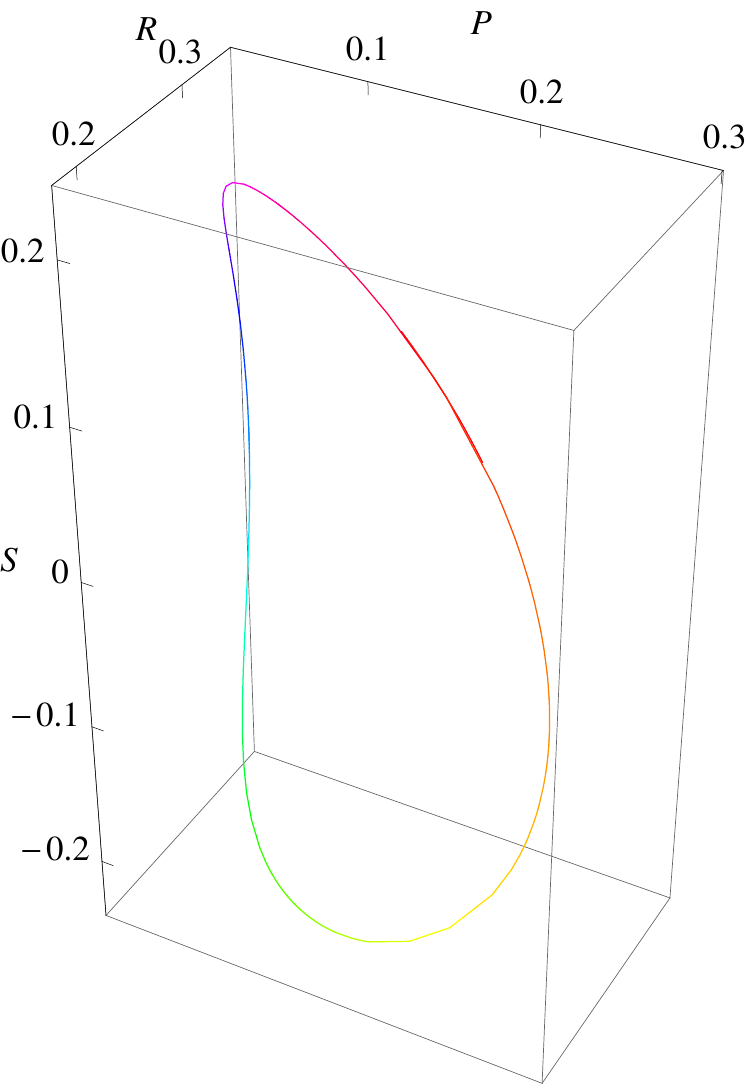}
\\
\raisebox{4.3cm}{\textbf{b}}~~
\includegraphics[scale=0.5]{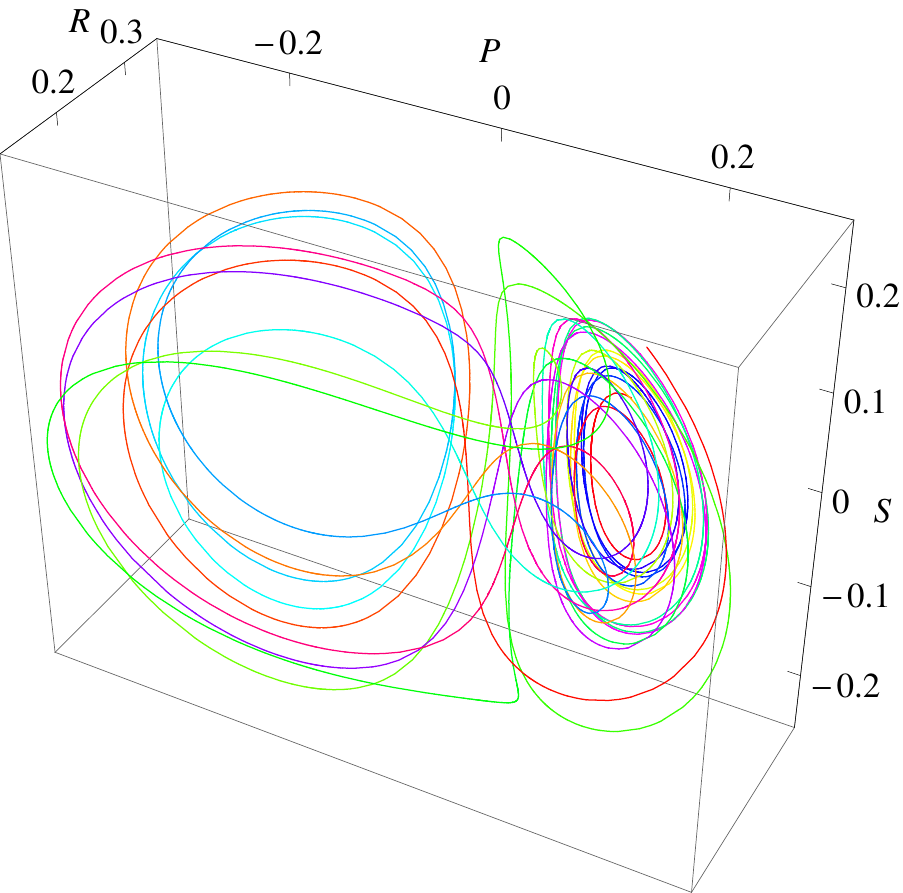}
\\
\raisebox{3cm}{\textbf{c}}~~
\includegraphics[scale=0.6]{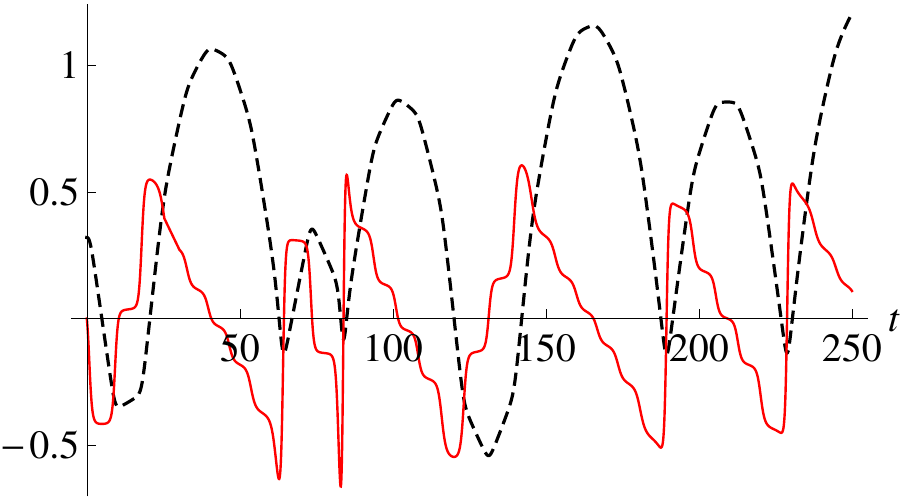}
%%%
% \raisebox{5cm}{\textbf{a}}~~
% \includegraphics[scale=0.5]{periodic-3D.pdf}
% ~~~~~~~
% \raisebox{5cm}{\textbf{b}}~~
% \includegraphics[scale=0.6]{chaotic-3D.pdf}
% \\ ~~ \\ 
% ~~~~~~~~~~~~~~~~~~~~~~~~~~~~~~~~~~~~~~~~~~~~
% \raisebox{3cm}{\textbf{c}}~~
% \includegraphics[scale=0.65]{chaotic-z3z4.pdf}
\caption{
\label{fig:3D}
(a) Unperturbed orbit in the $(P~R~S)$ space. 
%  (left), and its projection on the $(P~R)$ plane (right);
$\lambda=4, P(0)=R(0)=S(0)=0.2$.  
(b) Perturbed chaotic orbit in the $(P~R~S)$ subspace.
%  (left), and its projection on the $(P~R)$ plane (right);
$\lambda=4, P(0)=R(0)=S(0)=0.2$, and $\epsilon=2\times10^{-7}$.
(c) The evolution of $Z_3(t)\times10^3$ (black dotted) and $Z_4(t)\times10^{-3}$ (red solid).  
}
\end{center}
\end{figure}
%----------------------------------------------------------------------

   Let us examine the example
\begin{equation}
H(\tilde{\vect{Z}})  =   
\frac{1}{2}\left( {Z}_3^2\rme^{2\lambda{Z}_1} + \rme^{-2{Z}_1} + {Z}_2^2  \right)
+ \epsilon \, \frac{1}{2}\left( {Z}_3^2 + {Z}_4^2 \right) ,
\label{chaotic-H}
\end{equation}
where the $\epsilon$ term perturbs \eqref{MT-H(zeta)'}--\eqref{potential}.  The perturbation unfreezes 
the adiabatic invariant ${Z}_3=C$.  Physically this term represents an oscillation energy.

Figure \ref{fig:3D} shows typical solutions of the extended system \eqref{4D_Canonical-0}--\eqref{chaotic-H}.
In (a) and (b) we plot the orbits projected onto the 3-dimensional $(P~R~S)$ subspace.
%  (left) and on the 2-dimensional $(P~R)$ subspace (right), respectively.
Figure \ref{fig:3D} (a) depicts an  orbit of  the original integrable system: this   unperturbed case with $\epsilon=0$
shows the  orbit that is the  intersection of an energy sphere and a Casimir surface, as was seen   in 
Fig.~\ref{fig:leaves}.  Figure \ref{fig:3D} (b) illustrates a typical chaotic orbit that results from unfreezing the Casimir
$C = Z_3$ with   $\epsilon=2\times10^{-7}$, which then allows
the orbit to wander among different leaves (even into the negative $C$ domain).
In Fig.~\ref{fig:3D} (c),  $Z_3(t)$ is plotted  together with its conjugate variable $Z_4(t)$,
along with  the solution illustrated in (b).

Even after the canonization, the singularity ($S$-axis) of the original Lie-Poisson algebra
remains as a peculiar set around which the dynamics is dramatically modified by the singular perturbation
(here, the inclusion of the new variable $Z_4$ works as a singular perturbation, resulting in an increase of the number of degrees of freedom).
The perturbed system exhibits chaotic spin reversals, and 
chirality persists as long as the perturbation is weak enough for the orbit to make chaotic itinerancy among different leaves.

\section*{Acknowledgements}
ZY was supported by JSPS KAKENHI grant number 15K13532. 
PJM was supported by the  DOE Office of Fusion Energy Sciences, under DE-FG02-04ER- 54742.

\appendix
\section{Lie-Poisson brackets of 3-dimensional systems}

There is a systematic method for constructing Poisson brackets % of the form of (\ref{Poisson}) 
from any given Lie algebra.  
Such brackets are  called \emph{Lie-Poisson brackets}, because they were known to Lie in the 19th century.
Let $\mathfrak{g}$ be a Lie algebra with bracket $[~ , ~ ]$.
Take $\mathfrak{g}$ as the phase space and 
denote a linear functional on $\mathfrak{g}$ by $\langle \omega, ~ \rangle \in \mathfrak{g}^*$.
Choosing $\omega = \vect{X}$ we define,
for smooth functions $F(\vect{X})$ and $G(\vect{X})$,
\[
\{ F, G \} = \langle \vect{X}, [\partial_{\svect{X}} F , \partial_{\svect{X}} G ] \rangle,
\]
where $\partial_{\svect{X}} F$ is the gradient in $\mathfrak{g}$ of a function $F(\vect{X})$.
Because of  this construction the  Lie-Poisson bracket $\{ ~,~\}$ inherits 
bilinearity, anti-symmetry, and the Jacobi's identity from that of $[~,~ ]$. 

According to the Bianchi classification, the structure constants for 3-dimensional Lie
algebras have the form
\[
c^i_{jk} = \epsilon_{jks}m^{si} + \delta^{i}_{k} a_j -\delta^i_j a_k
\]
where $m$, a $3\times3$ symmetric matrix, and $a$, a triple, take on different values for the nine Lie
algebras as summarized in Table\,\ref{table:Bianch-3D}.
In the table
\[
\alpha = \left( \begin{array}{ccc}
0 & 1 & 0 \\
1 & 0 & 0 \\
0 & 0 & 0 \end{array} \right).
\]

%---------------------------------------------------------- TABLE Bianchi
\begin{table}[tb]
\caption{Bianchi classification of 3-dimensional Lie algebra
(after M.P. Ryan and L.C. Shepley\,\cite{Ryan-Sheply}).
% \textit{Homogeneous Relativistic Cosmologies},
% \textit{Princeton Ser. Phys.}
% (Princeton Univ. Press, Princeton, 1975), sec. 6.4).
}
\begin{center}
{\small
\begin{tabular}{l|l|l|l}
\hline
Class & Type  &  $m$  & $a_i$ 
\\ \hline \hline
A & $\mathrm{I}$ & $0$ & $0$
\\ \hline
A & $\mathrm{II}$ & $\mathrm{diag}(1,0,0)$ & $0$
\\ \hline
A & $\mathrm{VI}_{-1}$ & $-\alpha$ & $0$
\\ \hline
A & $\mathrm{VII}_{0}$ & $\mathrm{diag}(-1,-1,0)$ & $0$
\\ \hline
A & $\mathrm{VIII}$ & $\mathrm{diag}(-1,1,1)$ & $0$
\\ \hline
A & $\mathrm{IX}$ & $\mathrm{diag}(1,1,1)$ & $0$
\\ \hline \hline 
B & $\mathrm{III}$ & $-\frac{1}{2}\alpha$ & $-\frac{1}{2}\delta^i_3$
\\ \hline 
B & $\mathrm{IV}$ & $\mathrm{diag}(1,0,0)$ & $-\delta^i_3$
\\ \hline 
B & $\mathrm{V}$ & $0$ & $-\delta^i_3$
\\ \hline 
B & $\mathrm{VI}_{h\neq-1}$ & $\frac{1}{2}(h-1)\alpha$ & $-\frac{1}{2}\delta^i_3$
\\ \hline 
B & $\mathrm{VII}_{h=0}$ & $\mathrm{diag}(-1,-1,0)+\frac{1}{2}h\alpha$ & $-\frac{1}{2}h \delta^i_3$
\\ \hline 
\end{tabular}
}
\end{center}
\label{table:Bianch-3D}
\end{table}
%---------------------------------------------------------------------------------

The $3\times3$ antisymmetric matrices $J$ defined as (using lowered indices):
\[
J_{ij} = c^k_{ij} X_k 
\]
gives the Lie-Poisson brackets $\{F,G\} = \langle \partial_{\svect{X}} F, J \partial_{\svect{X}}G\rangle$;
cf. Table\,\ref{table:Bianchi-Casimirs} and \cite{Morrison1998}.
% P. J. Morrison, Rev. Mod. Phys. \textbf{70}, 467 (1998).
Notice that $J$ is linear with respect to $\vect{X}$.




\end{document}